\documentclass{article}
\usepackage{spconf,amsmath,graphicx,bm,amsfonts,slashbox}
\usepackage{cite}
\usepackage{algorithm}  
\usepackage{algorithmic}
\usepackage{caption}
\usepackage{epstopdf}
\usepackage{hyperref}
\usepackage{cleveref}
\usepackage[font = scriptsize]{subfig}
\usepackage{xfrac}

\newcommand{\R}{\mathbb{R}}
\newcommand{\mP}{\mathcal{P}}

\newcommand{\F}{\mathcal{F}}
\newcommand{\C}{\mathbb{C}}
\newcommand{\NoiProSeries}{\{\mathbf{y}_{i,\kappa}\in\R^L:|\kappa|\le K\}}

\DeclareMathOperator*{\argmin}{argmin}

\newcommand{\jane}[1]{{\color{red}{\bf\sf [Jane: #1]}}}
\newcommand{\m}{\mathrm{max}}

\title{Two-Dimensional Tomography from Noisy Projection Tilt Series taken at unknown view angles with Non-uniform Distribution}
\name{Lingda Wang, Zhizhen Zhao}
\address{Department of ECE and CSL, University of Illinois at Urbana-Champaign}
\begin{document}
%
\maketitle
\begin{abstract}
We consider a problem that recovers a 2-D object and the underlying view angle distribution from its noisy projection tilt series taken at unknown view angles. Traditional approaches rely on the estimation of the view angles of the projections, which do not scale well with the sample size and are sensitive to noise. We introduce a new approach using the moment features to simultaneously recover the underlying object and the distribution of view angles. This problem is formulated as constrained nonlinear least squares in terms of the truncated Fourier-Bessel expansion coefficients of the object and is solved by a new alternating direction method of multipliers (ADMM)-based algorithm. 
Our numerical experiments show that the new approach outperforms the expectation maximization (EM)-based maximum marginalized likelihood estimation in efficiency and accuracy. 
Furthermore, the hybrid method that uses EM to refine ADMM solution achieves the best performance. 
\end{abstract}
\section{Introduction}
\label{sec:1}
We consider the problem of reconstructing a 2-D image from multiple projection tilt series taken at unknown viewing angles. This problem is motivated by cryo-electron microscopy (cryo-EM) single particle reconstruction~\cite{frank2006three,cheng2018single,sorzano2015cryo} and cryo-electron tomography (cryo-ET)~\cite{wan2016cryo}. 
In both imaging modalities, projection images are taken at random unknown orientations. 

\begin{figure}[htb]
	\begin{center}
	\includegraphics[width=1 \linewidth]{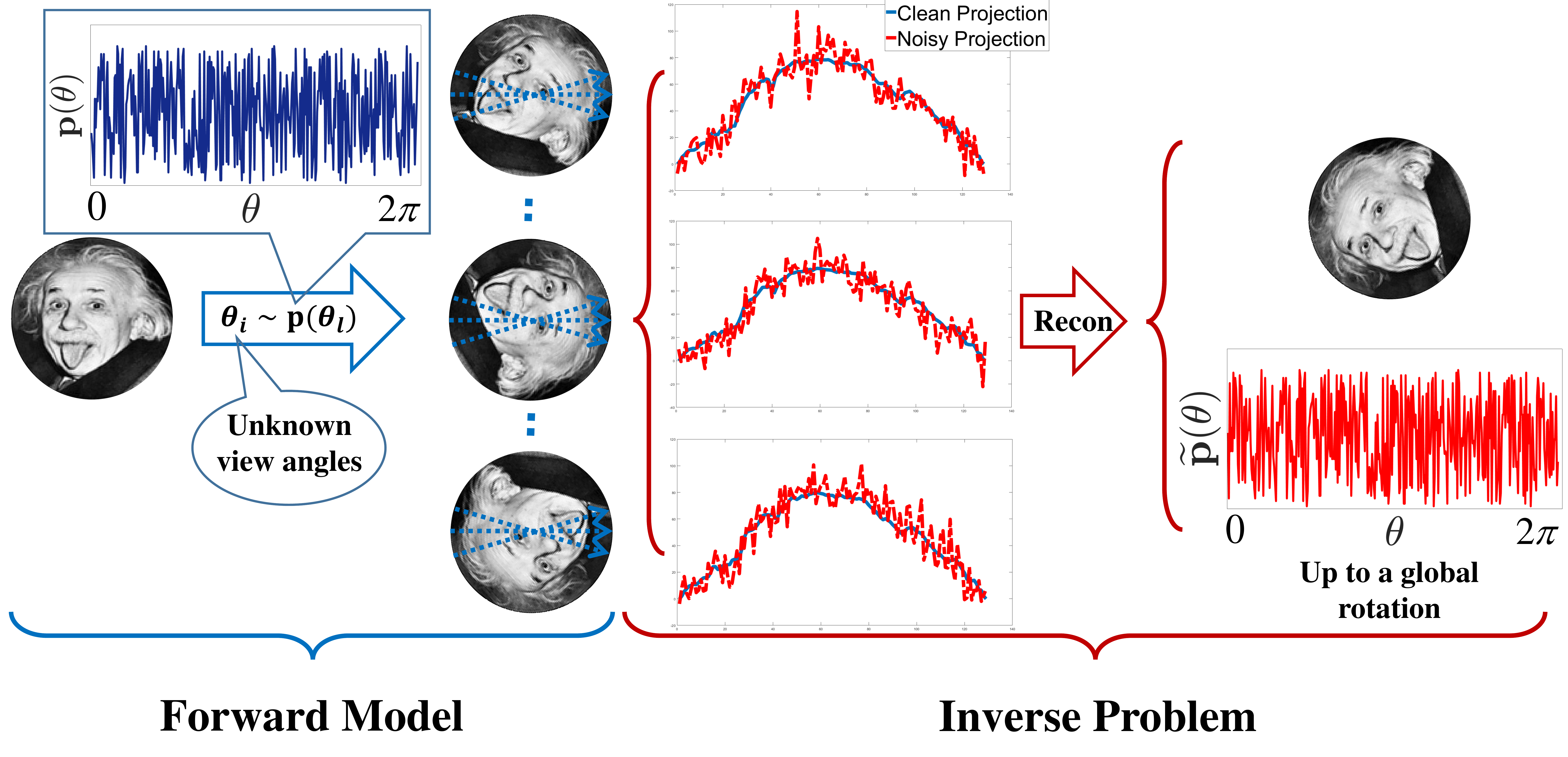}
	\end{center}
	\vspace{-0.6cm}
	\caption{Forward model: the projection tilt series are taken at unknown view angles of a 2-D object. It is equivalent to fixing the directions (blue dashed arrows) and rotating the object.  
	Our observed data are the noisy projection tilt series (red curves in the middle).
	The inverse problem: recover the 2-D object and $\mathbf{p}$ using those random tilt series.
	}
	\label{fig:explanation}
	\vspace{-0.3cm}
\end{figure}

We assume that our underlying 2-D object $f$ is a well behaved function in $\R^2$, i.e., $f\in\mathcal{V}\subset L_2(\R^2)\cap L_1(\R^2)$. The tilt angles are equally spaced with a small opening angle $\alpha$ and the number of tilts is $2K+1$ ($K \in \mathbb{Z}^+$) and satisfy $K\alpha \leq \frac{\pi}{3}$. 
The observation model for one tilt projection is,
\vspace{-0.2cm}
\begin{align}
    \label{eq:ob_model}
    \mathbf{y}_{i,\kappa} = \mathcal{P}_{\theta_i+\kappa\alpha}\left(f\right)+\mathbf{n}_{i,\kappa},
\end{align}
where $\mathbf{n}_{i,\kappa} \in \mathbb{R}^L$ are i.i.d Gaussian random vectors with zero mean and variance $\sigma^2$, and $\mathcal{P}_{\theta_i+\kappa\alpha}: \mathcal{V} \mapsto \mathbb{R}^L$ is a linear operator which takes the radon transform of the object at the angle $\theta_i+\kappa\alpha$ and samples the projection on $L$ equally spaced grid points. The view angles $\theta_i$ are drawn from $n_\theta$ equally spaced angles $\{ \phi_l = \frac{2\pi l}{n_\theta}\}_{l = 0}^{n_\theta-1}$ according to an unknown distribution $\mathbf{p}$.  
Given the data $\NoiProSeries_{i=1}^N$, 
our goal is to recover $f$ and the underlying view angle distribution $\mathbf{p}$ up to a global rotation as illustrated in Fig.~\ref{fig:explanation}. 

Most algorithms for 2-D tomography from unknown view angles rely on the estimation of the view angle ordering \cite{basu2000feasibility,basu2000uniqueness,coifman2008graph}. In particular,~\cite{coifman2008graph} uses diffusion maps~\cite{coifman2006diffusion} to estimate the view angles. Combining PCA, Wiener filtering, graph denoising and diffusion maps~\cite{singer2013two} is able 
to improve the performance of~\cite{coifman2008graph} on noisy projections. However, for extremely noisy data, it is impossible to accurately estimate the view angles and a large amount of data is required for image reconstruction. 

In this paper, instead of determining the view angle ordering of the projection tilts, we propose a new approach using moment features. Using invariant moment features for signal estimation was shown to be very efficient for large data in \textit{1-D multi-reference alignment}~\cite{bendory2017bispectrum,chen2018spectral, abbe2018multireference} and  
\textit{multi-segment reconstruction}~\cite{zehni2018multi}. The estimation of $f$ and $\mathbf{p}$ is formulated as a constrained weighted nonlinear least squares problem with moment features estimated from the projections. We introduce an alternating direction method of multipliers (ADMM) algorithm \cite{boyd2011distributed,lu65nonconvex} to solve the corresponding nonconvex optimization problem and empirically verify that the ADMM based algorithm is able to recover $f$ and $\mathbf{p}$ exactly up to a global rotation for the clean case. We approach the same problem using an expectation maximization (EM)-based algorithm for maximum marginalized likelihood estimation. The EM algorithm is typically sensitive to the initialization. Therefore, we compare the results between EM with random initialization and EM initialized the ADMM solution (ADMM + EM). 
Our numerical results demonstrate that ADMM and ADMM+EM are able to avoid getting stuck at bad local critical points and achieve more accurate reconstruction than EM with random initialization. 

\vspace{-0.2cm}
\section{Methods}
\vspace{-0.2cm}
\label{sec:2}
\subsection{Representation of the Image}
According to the Fourier slice theorem, the 1-D Fourier transform 
of the clean projection line, $\widehat{\mP_\theta\left(f\right)}$, corresponds to the central slice of the 2-D Fourier transform of the function, $\F\left(f\right)$, evaluated at angle $\theta$, namely,
\vspace{-0.2cm}
\begin{equation}
\label{eq:fst}
    \widehat{\mP_\theta\left(f\right)}(\xi)=\F\left(f\right)(\xi,\theta).
\end{equation}
For a bandlimited function $f$ with bandlimit $c$, its Fourier transform can be expanded on a set of orthonormal basis on a disc of radius $c$. Specifically, we use the FB basis defined as,
\vspace{-0.4cm}
\begin{equation}
\psi_c^{k, q}\left(\xi,\theta\right) =
\begin{cases}
N_{k, q}J_k\left( R_{k, q}\frac{\xi}{c} \right)e^{\imath k \theta} , & \xi \leq c,  \\
\hfil 0 ,  & \xi > c,
\end{cases}
\end{equation}
where $J_k$ is the $k^\text{th}$ order Bessel function of the first kind, $R_{k,q}$ is the $q^{\text{th}}$ root of $J_k$, and $N_{k,q}=\left(c\sqrt{\pi}\left|J_{k+1}\left(R_{k,q}\right)\right|\right)^{-1}$ is a normalization factor. 
Assuming that the object is also well concentrated in real domain, with the radius of support $R$, the Fourier transformed function can be well approximated by the truncated FB expansion~\cite{zhao2013fourier,zhao2016fast} according to the sampling criterion, $R_{k, q+1} \leq 2\pi c R$,
\vspace{-0.4cm}
\begin{align}
\label{eq:2}
\F\left(f\right)\left(\xi,\theta\right) \approx \sum_{k = -k_{\text{max}} }^{k_{\text{max}}} \sum_{q = 1}^{q_k} a_{k, q} \psi_c^{k, q}\left(\xi, \theta\right).
\end{align}
The truncation is important for removing spurious information from noise~\cite{klug1972three}. Combining~\eqref{eq:2} with~\eqref{eq:fst}, we get $\widehat{\mP_{\theta}\left(f\right)}[\xi] \approx \sum_{k = -k_\m}^{k_\m} \sum_{q = 1}^{q_k} a_{k, q} \psi_c^{k, q}(\xi, \theta):=\mathbf{\mathbf{\Psi}}_\theta \mathbf{a}[\xi]$,
where $\mathbf{\mathbf{\Psi}}_\theta$ contains $\psi_c^{k, q}(\xi, \theta)$ and $\mathbf{a}$ is the vectorized $a_{k,q}$'s.

We evaluate the Fourier coefficients $\widehat{\mathbf{y}}_{i, \kappa}[\xi_j]$ at non-equally spaced Gaussian quadrature points with the associated weights  $\{\xi_j, w_j \}_{j = 1}^{n_\xi}$using non-uniform FFT~\cite{dutt1993fast,fenn2007computation,barnett2018parallel},
\vspace{-0.2cm}
\begin{equation}
\label{eq:FT_y}
    \widehat{\mathbf{y}}_{i,\kappa}[\xi_j] = \widehat{\mP_{\theta_i+\kappa\alpha}\left(f\right)}[\xi_j]+\widehat{\mathbf{n}}_{i,\kappa}[\xi_j].
\end{equation}
The vectors $\widehat{\mathbf{y}}_i$ and $\widehat{\mathbf{n}}_i$ are vertical concatenations of the vectors $\widehat{\mathbf{y}}_{i, \kappa}$ and $\widehat{\mathbf{n}}_{i, \kappa}$ respectively. The population covariance matrix of $\widehat{\mathbf{n}}_{i}$ is a known block diagonal matrix, denoted by $\widehat{\bm \Sigma}$.

\vspace{-0.2cm}
\subsection{Method of Moments and ADMM}
\noindent \textbf{Moment Features: }
The entries of the first order moment from the clean tilt series $\bm{\mu}[j; \kappa] := \mathbb{E}_{\theta} \widehat{\mP_{\theta + \kappa \alpha}\left(f\right)}[\xi_j]$ can be expressed as a set of second order polynomials in terms of the expansion coefficients $\mathbf{a}$ and the view distribution $\mathbf{p}$,
\vspace{-0.4cm}
\begin{align}
\label{eq:1M}
\bm{\mu}\left[j;\kappa\right] & =\sum_{k = -k_\m}^{k_\m} \sum_{q = 1}^{q_k} \sum_{l = 0}^{n_\theta-1} \textstyle a_{k,q} \psi_c^{k,q}\left(\xi_j, \phi_l+\kappa\alpha\right)\mathbf{p}[l] \nonumber\\
& =  \sum_{k = -k_\m}^{k_\m} \sum_{q = 1}^{q_k} a_{k,q}\psi_c^{k,q}\left(\xi_j,\kappa\alpha\right)\widehat{\mathbf{p}}[-k]\nonumber\\
& = \mathbf{\Psi}\left(\mathbf{a}\circ \mathbf{g}(\widehat{\mathbf{p}})\right)\left[j;\kappa\right],
\end{align}
where $\widehat{\mathbf{p}}[k]$ is the Fourier coefficient of $\mathbf{p}$. The matrix $\mathbf{\Psi}$ contains $\psi_c^{k,q}\left(\xi_j,\kappa\alpha \right)$ with the rows indexed by $(j, \kappa)$ and the columns indexed by $(k, q)$. We use `$\circ$' to denote the entry-wise (Hardmard) product. The vector $\mathbf{g}(\widehat{\mathbf{p}})$ vertically concatenates $\widehat{\mathbf{p}}$.  Our second feature is the second order moment $\mathbf{C}$ of the projection tilt series, which provides a set of third order polynomials in terms of $\mathbf{a}$ and $\mathbf{p}$,
\vspace{-0.4cm}
\begin{align}
\label{eq:2M}
&\mathbf{C}[j_1;\kappa_1,j_2;\kappa_2] =
\sum_{k_1 = -k_\m}^{k_\m} \sum_{k_2 = -k_\m}^{k_\m} \sum_{q_1 = 1}^{q_{k_1}} \sum_{q_2 = 1}^{q_{k_2}} a_{k_1,q_1}\overline{a_{k_2,q_2}} \nonumber\\ 
&\quad\quad\times\psi_c^{k_1,q_1}\left(\xi_{j_1},\kappa_1\alpha\right)
\overline{\psi_c^{k_2,q_2}\left(\xi_{j_2},\kappa_2\alpha\right)}\widehat{\mathbf{p}}[k_2-k_1]\nonumber\\
& =\left(\mathbf{\Psi}\left(\mathbf{a}\mathbf{a}^*\circ\mathbf{H}(\widehat{\mathbf{p}})\right)\mathbf{\Psi}^* \right)[j_1;\kappa_1,j_2;\kappa_2],
\end{align}
where $\mathbf{C}\in\C^{(2K+1)n_\xi\times(2K+1)n_\xi}$ and $\mathbf{H}(\widehat{\mathbf{p}})$ is the matrix consists of $\widehat{\mathbf{p}}$. Assuming that the number of tilts is sufficiently large, i.e. $K \geq 1$, and that the distribution $\mathbf{p}$ is aperiodic, the first-order and second-order moments in~\eqref{eq:1M} and~\eqref{eq:2M} can uniquely determine the underlying object $\mathbf{a}$ and the view distribution $\mathbf{p}$, up to a global rotation.

We use $\widehat{\mathbf{y}}_i$ to construct the empirical unbiased estimators of $\bm{\mu}$ and $\mathbf{C}$, that is 
\vspace{-0.4cm}
\begin{align}
\widetilde{\bm{\mu}} = \frac{1}{N}\sum_{i=1}^N\widehat{\mathbf{y}}_i,\quad 
\widetilde{\mathbf{C}}=\frac{1}{N}\sum_{i=1}^N \widehat{\mathbf{y}}_i\widehat{\mathbf{y}}_i^*-\widehat{\mathbf{\Sigma}}.
\label{eq:sample_estimates}
\end{align}
As the sample size $N \rightarrow \infty$, the sample estimates in~\eqref{eq:sample_estimates} converge to $\bm{\mu}$ and $\mathbf{C}$. Therefore, we formulate the reconstruction from the moment features as the following constrained weighted nonlinear least squares, 
\vspace{-0.3cm}
\begin{align}
\label{eq:nonconvex}
\left(\widetilde{\mathbf{a}},\,\widetilde{\mathbf{p}}\right) = &\argmin_{\mathbf{a},\mathbf{p}}\,
\begin{aligned}
& \textstyle \frac{\lambda_1}{2}\|\mathbf{\Psi}\left(\mathbf{a}\circ \mathbf{g}(\widehat{\mathbf{p}})\right)-\widetilde{\bm{\mu}}\|_w^2 \\
&\textstyle +\frac{\lambda_2}{2}\|\mathbf{\Psi}\left(\mathbf{a}\mathbf{a}^*\circ\mathbf{H}(\widehat{\mathbf{p}})\right)\mathbf{\Psi}^*-\widetilde{\mathbf{C}}\|_{W}^2,
\end{aligned} \nonumber\\
&\text{s.t. } \quad \mathbf{p} \ge \mathbf{0} \quad \text{ and } \quad  \mathbf{1}^\top \mathbf{p}=1,
\end{align}
where $\lambda_1 > 0$ and $\lambda_2 > 0$ are weights corresponding to the first and second moment, respectively, and we use $\mathbf{1}$ to represent an all ones column vector. The corresponding norms $\|\cdot\|_w$ and $\|\cdot\|_{W}$ are weighted $\ell_2$ and Frobenius norms with $\|\bm{\mu}\|_{w}^2 = \sum_{j,\kappa}\bm{\mu}\left[j;\kappa\right]^2\xi_j w_j$ and $\|\mathbf{C}\|_{W}^2 = \sum_{j_1,j_2,\kappa_1,\kappa_2}\mathbf{C}\left[j_1;\kappa_1,j_2;\kappa_2\right]^2\xi_{j_1}\xi_{j_2}w_{j_1}w_{j_2}$.

\noindent \textbf{An ADMM Approach: }
We rewrite the problem in~\eqref{eq:nonconvex} into 
\vspace{-0.3cm}
\begin{align}
\label{cost_func:ADMM}
&\min_{\mathbf{a},\mathbf{z},\mathbf{p}} \,
\begin{aligned}
&\textstyle \frac{\lambda_1}{2}\|\mathbf{\Psi}\left(\mathbf{a}\circ \mathbf{g}(\widehat{\mathbf{p}})\right)-\widetilde{\bm{\mu}}\|_w^2+ \frac{\lambda_1}{2}\|\mathbf{\Psi}\left(\mathbf{z}\circ \mathbf{g}(\widehat{\mathbf{p}})\right) -\widetilde{\bm{\mu}}\|_w^2 \\
&\textstyle  + \frac{\lambda_2}{2}\|\mathbf{\Psi}\left(\mathbf{a}\mathbf{z}^*\circ\mathbf{H}(\widehat{\mathbf{p}})\right)\mathbf{\Psi}^*-\widetilde{\mathbf{C}}\|_{W}^2 
\end{aligned}  \nonumber  \\
& \textup{s.t.} \quad \, \mathbf{a} = \mathbf{z}, \quad   \mathbf{p} \ge \mathbf{0}, \quad \text{and} \quad  \mathbf{1}^\top \mathbf{p}=1.
\end{align}
We fix $\widehat{\mathbf{p}}[0] = 1$, which ensures that the sum of all entries of $\mathbf{p}$ is 1, and relax the condition that $\mathbf{p}$ is non-negative. We then construct the corresponding augmented Lagrangian, 
\begin{align}
\label{Lag:1}
&\mathcal{L}(\mathbf{a},\mathbf{z},\mathbf{p};\mathbf{s}) = \textstyle \frac{\lambda_1}{2} \left \|\mathbf{\Psi}\left(\mathbf{a}\circ \mathbf{g}(\widehat{\mathbf{p}})\right)-\widetilde{\bm{\mu}} \right \|_w^2+\frac{\lambda_1}{2} \left \|\mathbf{\Psi}\left(\mathbf{z}\circ \mathbf{g}(\widehat{\mathbf{p}})\right) \right. \nonumber\\
&\textstyle \left.-\widetilde{\bm{\mu}} \right \|_w^2+\frac{\lambda_2}{2}\left \|\mathbf{\Psi}\left(\mathbf{a}\mathbf{z}^*\circ\mathbf{H}(\widehat{\mathbf{p}})\right)\mathbf{\Psi}^*-\widetilde{\mathbf{C}} \right \|_{W}^2 +\frac{\rho}{2}\left \|\mathbf{a}-\mathbf{z} + \mathbf{s}\right \|_2^2,
\end{align}
where $\mathbf{s}$ is the dual variable and the value of the penalty parameter $\rho>0$ will be discussed in Sec.~\ref{sec:3}. The ADMM algorithm alternates between the primal updates of variables $\mathbf{a}$, $\mathbf{z}$, and $\mathbf{p}$, and the dual update for $\mathbf{s}$.
\vspace{-0.2cm}
\begin{algorithm}[htb]
	\caption{The ADMM Algorithm}
	\label{algo:ADMM}
	\begin{algorithmic}[1]
		\STATE {\bfseries Input:} The sample estimates of the first- and second-order moments $\widetilde{\bm{\mu}}$ and $\widetilde{\mathbf{C}}$ in \eqref{cost_func:ADMM}
		\STATE {\bfseries Output:}  $\widetilde{\mathbf{a}},\widetilde{\mathbf{p}}$ 
		\STATE {\bfseries Initialization:} random initialization of  $\mathbf{a}^{(0)},\mathbf{z}^{(0)},\mathbf{p}^{(0)}$, and $\mathbf{s}^{(0)} = \mathbf{0}$ 
		\REPEAT
		\STATE  $\mathbf{a}^{(t+1)}=\argmin_{\mathbf{a}}\,\mathcal{L}(\mathbf{a},\mathbf{z}^{(t)},\mathbf{p}^{(t)};\mathbf{s}^{(t)})$
		\STATE  $\mathbf{z}^{(t+1)}=\argmin_{\mathbf{z}}\,\mathcal{L}(\mathbf{a}^{(t+1)},\mathbf{z},\mathbf{p}^{(t)};\mathbf{s}^{(t)})$
		\STATE $\mathbf{\mathbf{p}}^{(t+1)}=\argmin_{\mathbf{p}}\,\mathcal{L}(\mathbf{a}^{(t+1)},\mathbf{z}^{(t+1)},\mathbf{p};\mathbf{s}^{(t)})$
		\STATE $\mathbf{s}^{(t+1)}=\mathbf{s}^{(t)} +\mathbf{a}^{(t+1)}-\mathbf{z}^{(t+1)}$
		\UNTIL{convergence}
		\RETURN $\mathbf{a}^{(t)},\mathbf{p}^{(t)}$
	\end{algorithmic}
\end{algorithm}

\vspace{-0.4cm}
We introduce a diagonal matrix $D_w$, with $D_w[j, j] = \sqrt{w_j\xi_j}$ and replace  $\bm \Psi$, $\widetilde{\bm\mu}$,  and $\widetilde{\mathbf{C}}$ with $\bm{\Psi}_w = D_w \bm{\Psi} $, $\widetilde{\bm\mu}_w = D_w\widetilde{\bm\mu}$, and $\widetilde{\mathbf{C}}_w = D_w\widetilde{\mathbf{C}}D_w^*$. In this way, the weighted norms in~\eqref{Lag:1} can be replaced by unweighted norms. The introduction of $\mathbf{z}$ allows the primal update to be split into three linear problems. 
Specifically, given $\mathbf{z}$ and $\widehat{\mathbf{p}}$, we can express $\mathbf{\Psi}_w\left(\mathbf{a}\circ \mathbf{g}(\widehat{\mathbf{p}})\right)$ and $\mathbf{\Psi}_w\left(\mathbf{a}\mathbf{z}^*\circ\mathbf{H}(\widehat{\mathbf{p}})\right)\mathbf{\Psi}_w^*$ in terms of linear transform of $\mathbf{a}$, i.e. $\mathbf{A}_1(\widehat{\mathbf{p}})\mathbf{a}$ and $\mathbf{A}_2(\mathbf{z},\widehat{\mathbf{p}})\mathbf{a}$, respectively. Therefore, the subproblem of updating $\mathbf{a}$ is given by
\vspace{-0.2cm}
\begin{align}
\label{argmin:1}
& \mathbf{a}^{(t+1)} = \argmin_{\mathbf{a}} \textstyle \frac{\lambda_1}{2}\|\mathbf{A}^{(t)}_1\mathbf{a}-\widetilde{\bm{\mu}}_w\|_2^2 + \frac{\lambda_2}{2}\|\mathbf{A}^{(t)}_2 \mathbf{a}-\widetilde{\bm{c}}_w\|_2^2   \nonumber \\
& \textstyle \quad \quad \quad \quad \quad \quad + \frac{\rho}{2}\|\mathbf{a}-\mathbf{z}^{(t)} + \mathbf{s}^{(t)}\|_2^2 \nonumber \\
&\textstyle = \left[\lambda_1\left(\mathbf{A}^{(t)}_1\right)^*\mathbf{A}^{(t)}_1+\lambda_2 \left( \mathbf{A}^{(t)}_2\right)^*\mathbf{A}^{(t)}_2+\rho\mathbf{I} \right]^{-1}\left [\lambda_1 \left(\mathbf{A}^{(t)}_1\right)^*\right. \nonumber \\
& \quad \quad  \left. \times \widetilde{\bm{\mu}}_w  +\lambda_2 \left(\mathbf{A}^{(t)}_2\right)^*\widetilde{\bm{c}}_w+\rho(\mathbf{z}^{(t)} - \mathbf{s}^{(t)}) \right ],
\vspace{-0.2cm}
\end{align}
where $\widetilde{\bm{c}}_w$ is the vectorized $\widetilde{\mathbf{C}}_w$. Similarly, the variables $\mathbf{z}$ and $\mathbf{p}$ are updated by solving systems of linear equations. 
The steps of our algorithm are detailed in Alg.~\ref{algo:ADMM}. With the estimated coefficients $\mathbf{a}$, we use the analytical expression of the inverse Fourier transform of $\psi_c^{k, q}$ to reconstruct the image in the real domain as detailed in~\cite{zhao2016fast}. The computational complexity is $\mathcal{O}(N L \log L)$ for non-uniform FFT \eqref{eq:FT_y}, $\mathcal{O}(N L^2)$ moment features generation \eqref{eq:sample_estimates}, and $\mathcal{O}(L^6)$ for each iteration in Alg.~\ref{algo:ADMM}. 

\subsection{Maximizing the Marginalized Likelihood via EM}
The observation model in~\eqref{eq:ob_model} implies that $\widehat{\mathbf{y}}_i$ can be viewed as a sample randomly drawn from a set of $n_\theta$ Gaussian distributions. The marginalized likelihood of $\widehat{\mathbf{y}}_i$ given $(\mathbf{a}, \mathbf{p})$ is,
\vspace{-0.4cm}
\begin{equation}
\label{eq:likelihood}
P\left(\widehat{\mathbf{y}}_i|\mathbf{a}, \mathbf{p}\right) = \frac{1}{Z} \sum_{l = 0}^{n_\theta -1} \mathbf{p}[l] e^{-\frac{1}{2}(\widehat{\mathbf{y}}_i-\mathbf{\Psi}_{l}\mathbf{a})^*\widehat{\mathbf{\Sigma}}^{-1}(\widehat{\mathbf{y}}_i-\mathbf{\Psi}_{l}\mathbf{a})}, \nonumber
\end{equation}
where $\mathbf{\Psi}_{l} = \left[ 
\mathbf{\Psi}_{\phi_l-K\alpha},
\cdots,
\mathbf{\Psi}_{\phi_l}, 
\cdots,
\mathbf{\Psi}_{\phi_l+K\alpha}
\right]^\top$ and $Z$ is the normalization factor. 

Then the maximum marginalized log-likelihood estimation is formulated as,
\vspace{-0.3cm}
\begin{align}
\label{eq:MLE}
\max_{\mathbf{a},\,\mathbf{p}}\,\sum_{i=1}^N\ln P\left(\widehat{\mathbf{y}}_i|\mathbf{a}, \mathbf{p} \right)\, \text{ s.t. }\,  \mathbf{p} \ge \mathbf{0} \,\, \text{ and } \,\,  \mathbf{1}^\top \mathbf{p}=1.
\end{align}
We apply EM to solve~\eqref{eq:MLE} and iterate between the expectation step (E-step), 
\vspace{-0.4cm}
\begin{align}
\mathbf{\pi}^{(t)}\left(\phi_l|\widehat{\mathbf{y}}_i\right)=\frac{\mathbf{p}^{(t)}[l]e^{-\frac{1}{2}(\widehat{\mathbf{y}}_i-\mathbf{\Psi}_{l}\mathbf{a}^{(t)})^*\widehat{\mathbf{\Sigma}}^{-1}(\widehat{\mathbf{y}}_i-\mathbf{\Psi}_{l}\mathbf{a}^{(t)})}}{\sum_{l=0}^{n_\theta-1}\mathbf{p}^{(t)}[l]e^{-\frac{1}{2}(\widehat{\mathbf{y}}_i-\mathbf{\Psi}_{l}\mathbf{a}^{(t)})^*\widehat{\mathbf{\Sigma}}^{-1}(\widehat{\mathbf{y}}_i-\mathbf{\Psi}_{l}\mathbf{a}^{(t)})}}, \nonumber
\end{align}
and the maximization step (M-step),
\vspace{-0.3cm}
\begin{align}
&\mathbf{p}^{(t+1)}[l] =\frac{1}{N}\sum_{i=1}^N\mathbf{\pi}^{(t)}(\phi_l|\widehat{\mathbf{y}}_i), \nonumber \\
&\mathbf{a}^{(t+1)}  =
\left[\sum_{i=1}^N\sum_{l=0}^{n_\theta-1}\mathbf{\Psi}_{l}^*\widehat{\mathbf{\Sigma}}^{-1}\mathbf{\Psi}_{l}\mathbf{\pi}^{(t)}\left(\phi_l|\widehat{\mathbf{y}}_i\right)\right]^{\dagger} \nonumber \\
& \quad \quad \quad \quad \times \left[\sum_{i=1}^N\sum_{l=0}^{n_\theta-1}\mathbf{\Psi}_{l}^*\widehat{\mathbf{\Sigma}}^{-1}\mathbf{y}_i\mathbf{\pi}^{(t)}\left(\phi_l|\widehat{\mathbf{y}}_i\right)\right].
\end{align}
We can choose a random initialization for $\mathbf{a}^{0}$ and $\mathbf{p^{0}}$ or use the results from Alg.~\ref{algo:ADMM} as the initial points. The latter is a combined method called ADMM + EM. The computational complexity is made up with $\mathcal{O}(NL\log L)$ for non-uniform FFT \eqref{eq:FT_y} and $\mathcal{O}(N L^4+L^6)$ for each EM iteration.

\vspace{-0.2cm}
\section{Numerical Experiments}
\label{sec:3}
\vspace{-0.2cm}

We apply the algorithms described above (ADMM, EM, and ADMM + EM) to simulated projection tilts and evaluate their performances for image reconstruction and the recovery of the viewing angle distribution. The results are determined up to a global rotation. Our numerical experiments are run in MATLAB on a computer with an Intel i7-4770 CPU.

\begin{figure}[t!]
\captionsetup[subfloat]{farskip=2pt,captionskip=1pt}
	\centering 
    \subfloat[Original]{
		\includegraphics[width=0.22\linewidth]{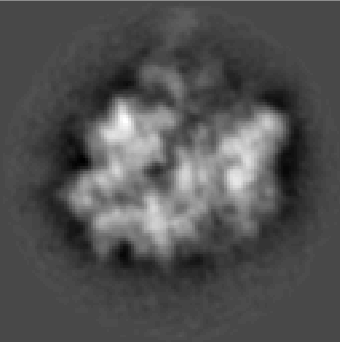}
		\label{fig:clean_ori}
		}
	\subfloat[Reconstructed]{
		\includegraphics[width=0.22\linewidth]{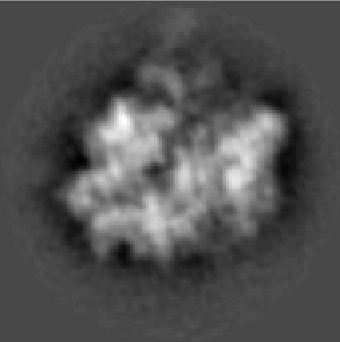}
		\label{fig:clean_recon}
		}
	\subfloat[ADMM, $\widetilde{\mathbf{p}}(\theta)$]{
		\includegraphics[width=0.42\linewidth]{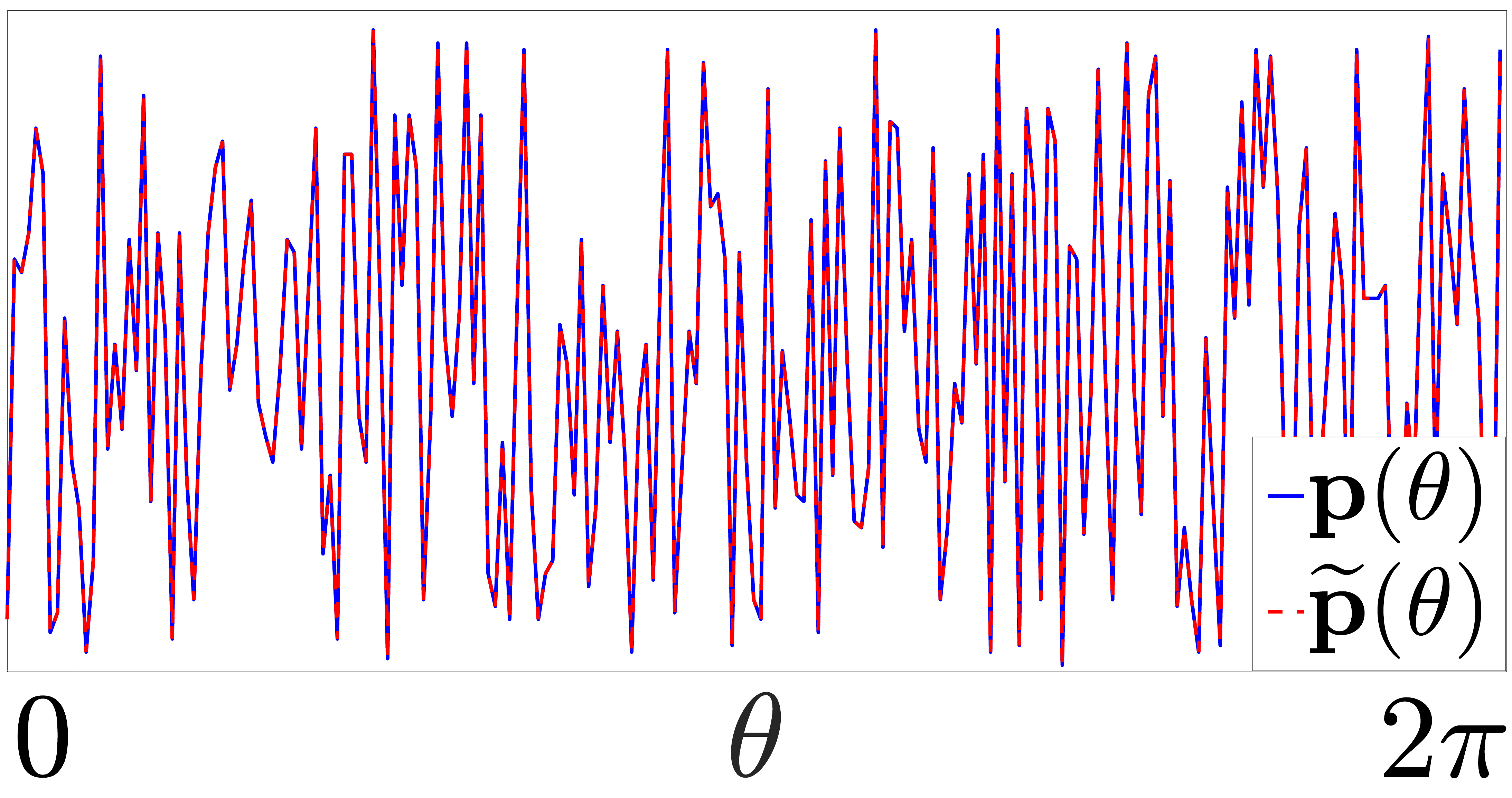}
		\label{fig:clean_pmf}
        }
    \vspace{-0.3cm}
	\caption{ADMM performance on clean random projection tilt series of a ribosome 70S image: \protect \subref{fig:clean_ori} Original image, \protect \subref{fig:clean_recon} recovered image, \protect \subref{fig:clean_pmf} true and estimated view distributions.}
	\label{fig:clean}
	\vspace{-0.5cm}
\end{figure}

\begin{figure}[t!]
\vspace{-0.5cm}
\captionsetup[subfloat]{farskip=2pt,captionskip=1pt}
	\centering 
	\subfloat[Original]{
		\includegraphics[width=0.23\columnwidth]{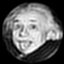}
		\label{fig:ein1}
		}
	\subfloat[ADMM]{
		\includegraphics[width=0.23\columnwidth]{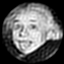}
		\label{fig:ein2}
		}
	\subfloat[EM]{
		\includegraphics[width=0.23\columnwidth]{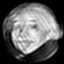}
		\label{fig:ein3}
		}
	\subfloat[ADMM+EM]{
		\includegraphics[width=0.23\columnwidth]{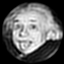}
		\label{fig:ein4}
	 }\\
	\subfloat[ADMM, $\widetilde{\mathbf{p}}(\theta)$]{
		\includegraphics[width=0.32\linewidth]{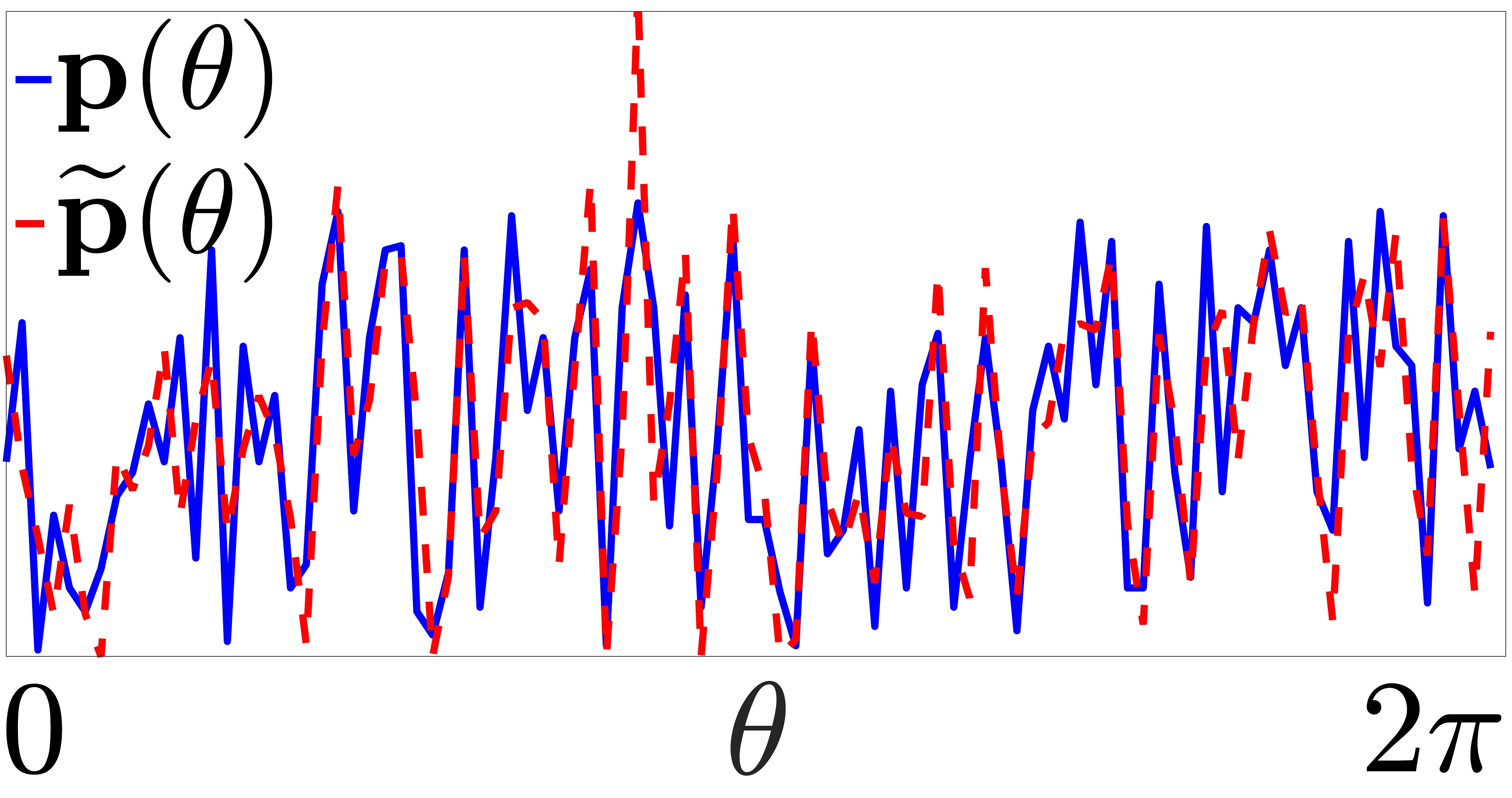}
		\label{fig:ein5}
		}
	\subfloat[EM, $\widetilde{\mathbf{p}}(\theta)$]{
		\includegraphics[width=0.32\linewidth]{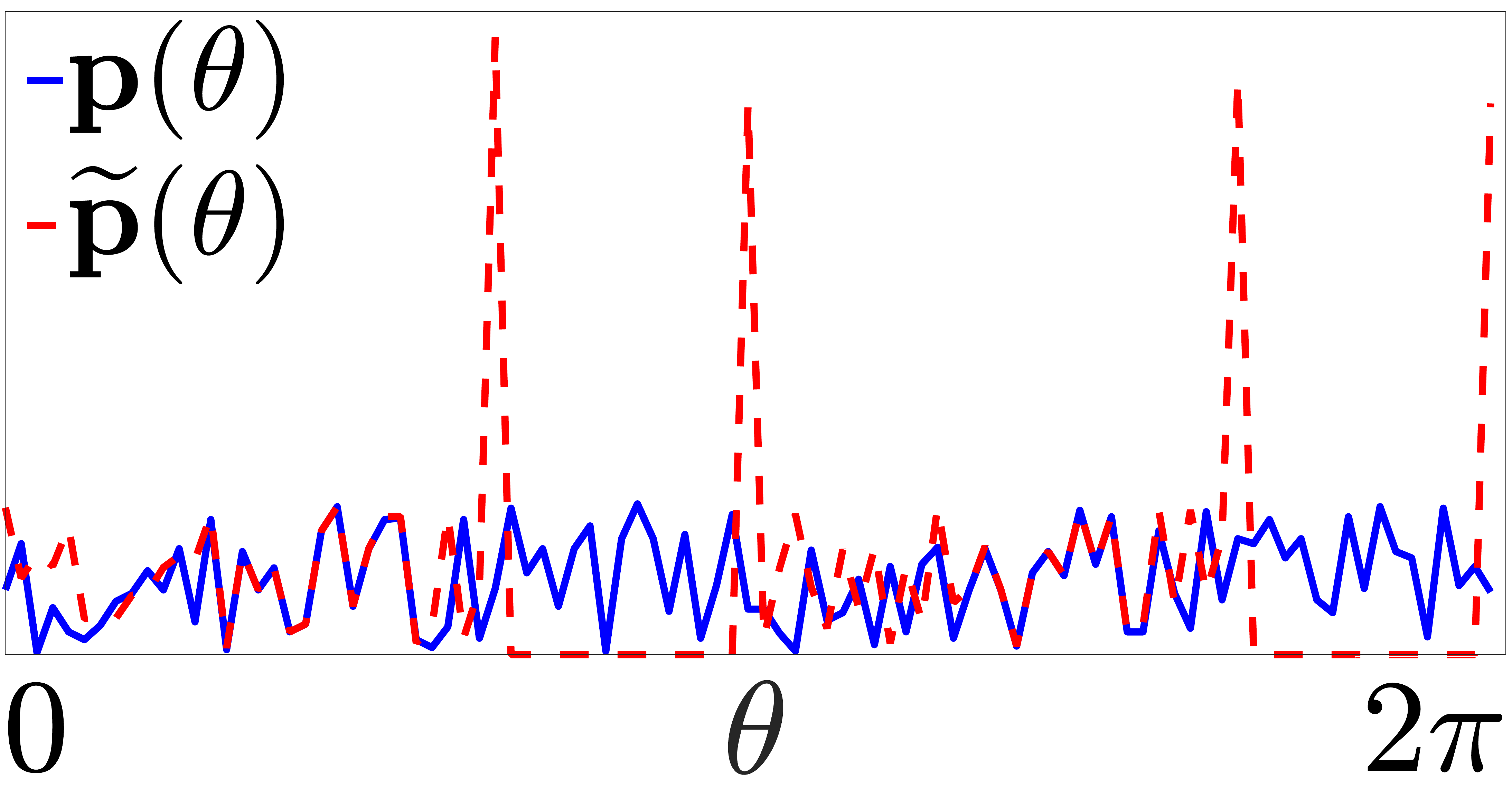}
		\label{fig:ein6}
		}
	\subfloat[ADMM+EM, $\widetilde{\mathbf{p}}(\theta)$]{
		\includegraphics[width=0.32\linewidth]{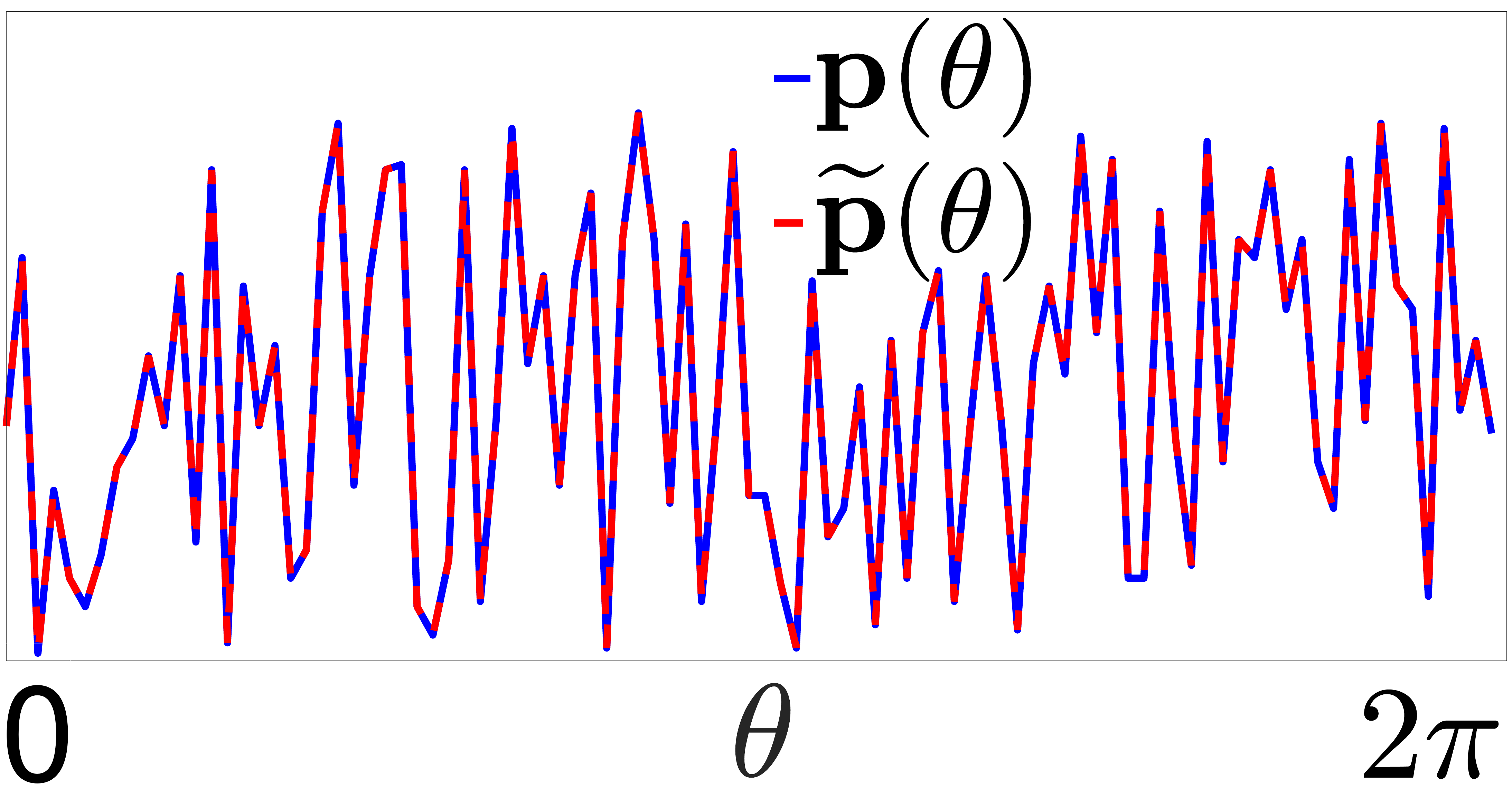}
		\label{fig:ein7}
	}
		\vspace{-0.4cm}
	\caption{Recovering the image of Einstein from projection tilt series: \protect \subref{fig:ein1} Original image; \protect \subref{fig:ein2} \protect \subref{fig:ein5}  ADMM, RE=4.48\%; \protect \subref{fig:ein3} \protect \subref{fig:ein6}  EM, RE=16.98\%; \protect \subref{fig:ein4} \protect \subref{fig:ein7} ADMM+EM, RE=0.82\%.}
	\label{fig:einstein_result}
	\vspace{-0.6cm}
\end{figure}

We define the signal-to-noise ratio (SNR), relative error (RE) and total variation (TV) by
\vspace{-0.4cm}
\begin{align*}
&\text{SNR [dB]} = 10\log_{10}(\text{Var}(\mathbf{y}_{i,\kappa})/\sigma^2),\\
&\text{RE}(I,\widetilde{I}) = \min_{\theta \in [0, 2\pi)}\|\mathcal{R}_{\theta}(\widetilde{I})-I\|_{F}/\|I\|_{F},
\end{align*}
\begin{align*}
\text{TV}(\mathbf{p},\widetilde{\mathbf{p}})=\min_{l\in\{0,n_{\theta}-1\}}\|\mathcal{S}_l(\widetilde{\mathbf{p}})-\mathbf{p}\|_1, \nonumber   
\end{align*}
where the digital image $I$ takes samples of $f$ on the Cartesian grid and $\widetilde{I}$ is the recovery; $\mathcal{R}_{\theta}(\cdot)$ is an operator that rotates the image by $\theta$ and $\mathcal{S}_l$ is an operator that the cyclic shift $\widetilde{\mathbf{p}}$ by $l$ positions.
\begin{figure}[h!]
\vspace{-0.2cm}
	\centering
	\includegraphics[scale=0.21]{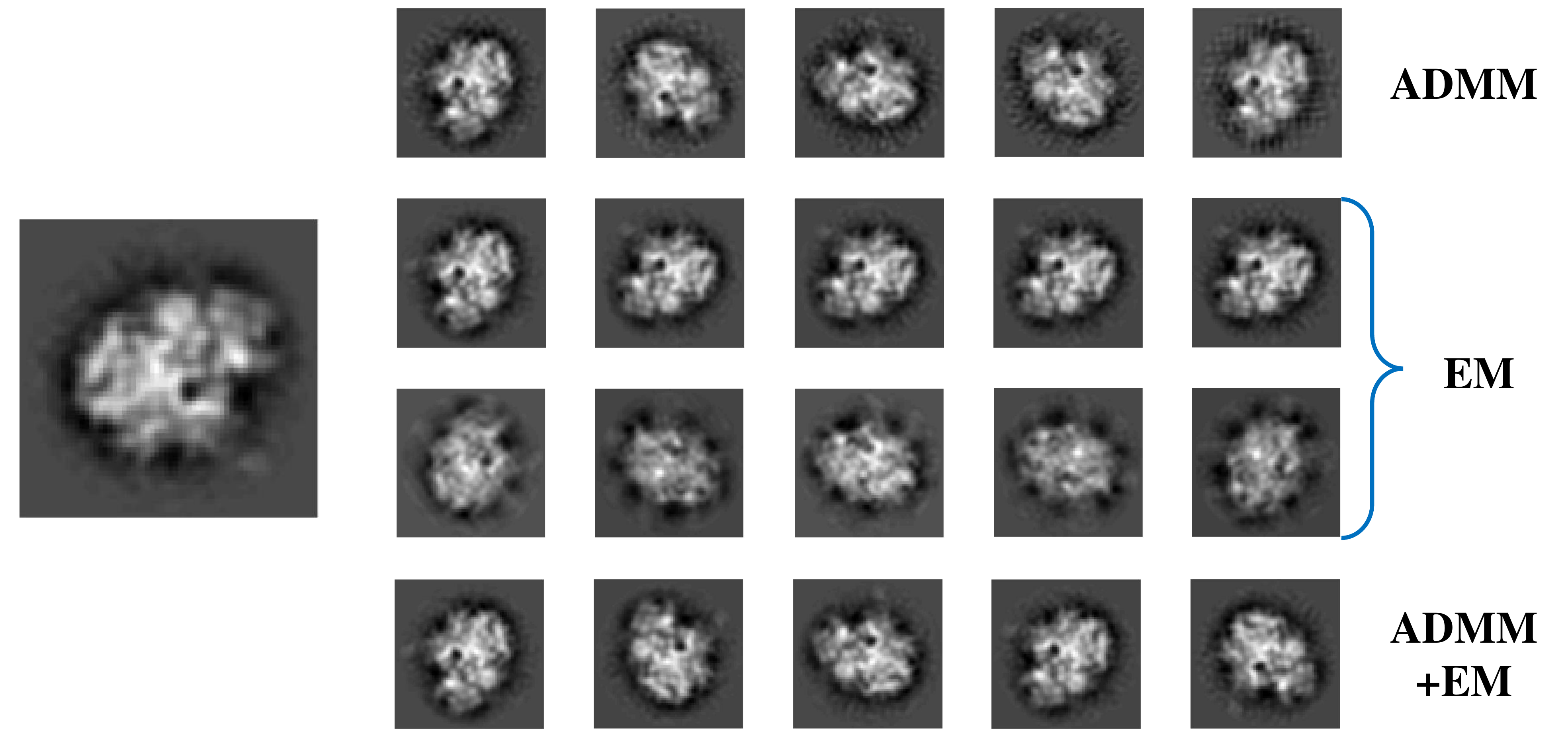}
		\vspace{-0.4cm}
	\caption{Comparisons of the reconstruction using ADMM, EM and ADMM+EM. \textit{The first column:} original image, and \textit{the second to sixth columns}: samples of the recovered images from noisy projection tilts with $\text{SNR [dB]} =$ 6.61, -0.32, -4.38, -7.25, -9.49, respectively. The results in the second row show the good recoveries while the third row shows the results goes into spurious local critical points of EM. }
	\label{fig:general_result}
	\vspace{-0.3cm}
\end{figure}

We first test the ADMM algorithm on recovering a projection image of 70S ribosome (see Fig.~\ref{fig:clean_ori}) from clean random tilt series. The size of the image is $128 \times 128$ and we generated $N = 10^4$ clean tilt series with $13$ equally spaced projection lines, where the angle between two neighboring lines is $1.5\mathrm{deg}$. The estimated bandlimit is 0.3 and we set the parameters in~\eqref{Lag:1} to be $\lambda_1=1$, $\lambda_2=0.5$, and $\rho=1$.  Our algorithm is able to achieve exact recovery with $\text{RE} \leq 10^{-5}$ and a perfectly matching viewing angle distributiocn (see Fig.~\ref{fig:clean}).

\begin{table}[htb]
    \vspace{-0.3cm}
	\caption{Success Rate of ADMM, EM and ADMM+EM}
	\centering
	\vspace{-0.4cm}
	\begin{tabular}{| c || r|r|r|r|r |}
		\hline \hline
		\backslashbox{Alg.}{SNR} &  6.61  &-0.32 &-4.38 &-7.25 &-9.49 \\ [0.5ex]
		\hline		
		ADMM & 100\% & 100\% & 100\% & 100\% & 100\%\\
		EM  & 35\% & 50\% & 45\% & 50\% & 40\% \\
		ADMM+EM & 100\% & 100\% & 100\% & 100\% & 100\% \\[0.5ex]
		\hline
	\end{tabular}
	\label{table:success_rate}
\end{table}

In the second experiment, we recover the picture of Einstein of size $64\times 64$ pixels shown in Fig.~\ref{fig:ein1} from simulated noisy random tilt series. We generated $N = 10^4$ random tilt series that contain 13 equally spaced projection lines ($K = 6$ and $\alpha=3.8\text{deg}$). The noise variance $\sigma^2 = 10$ (SNR$= 17.46$dB) and the estimated band limit is $0.3$. The parameters in~\eqref{Lag:1} are chosen to be $\lambda_1=1,\,\lambda_2=0.01$, and $\rho=1$. Fig.~\ref{fig:einstein_result} shows that compared to EM, ADMM and ADMM+EM exhibit advantages both in the image recovery and $\mathbf{p}$ estimation as EM easily gets stuck at bad local optimal with views concentrated on only a few directions. 

\begin{figure}[t!]
\captionsetup[subfloat]{farskip=2pt,captionskip=1pt}
	\centering 
    \subfloat[Image Reconstruction]{
		\includegraphics[width=0.48\linewidth]{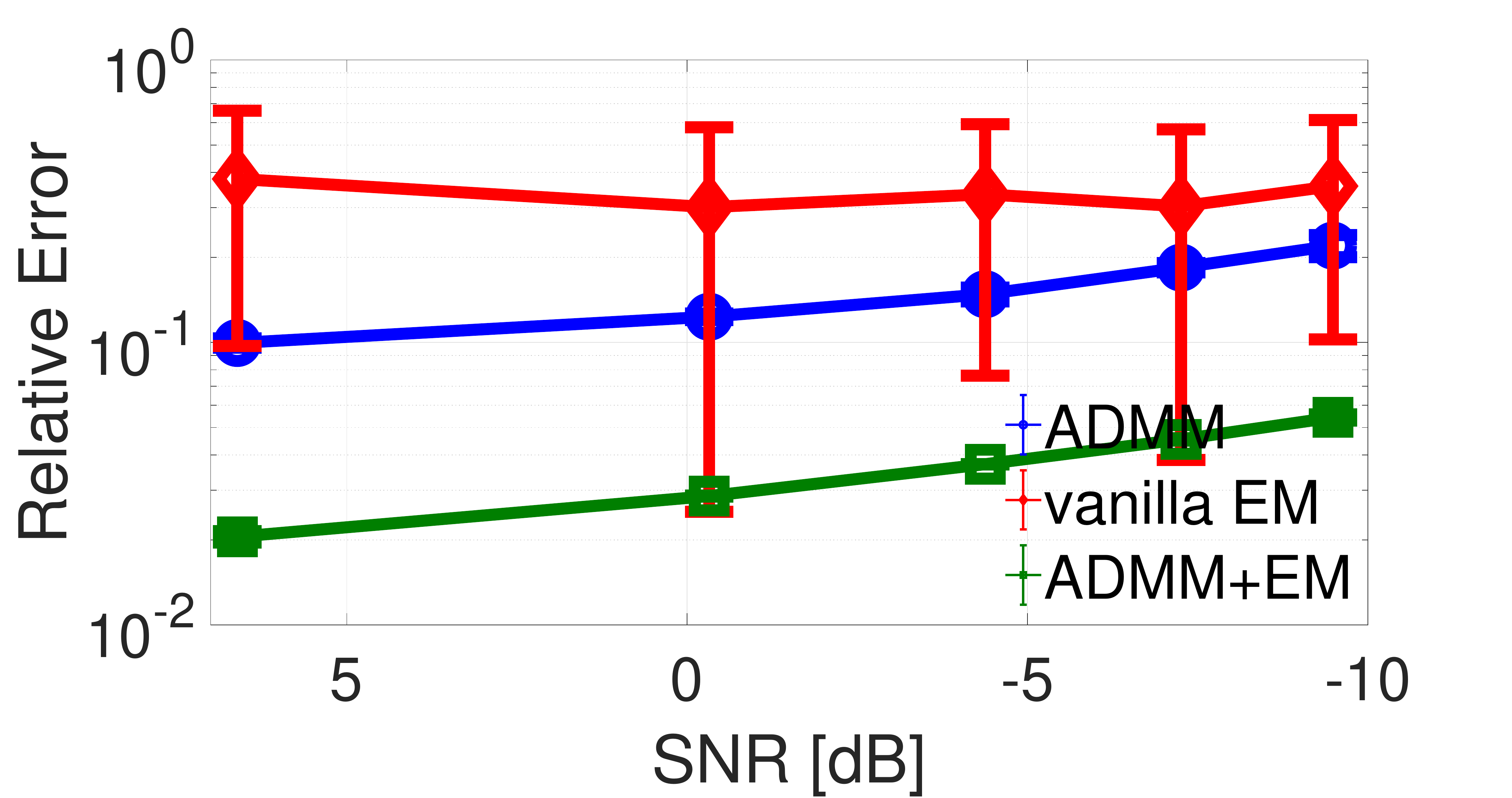}
		\label{fig:RE}
		}
    \subfloat[View Distribution]{
		\includegraphics[width= 0.48\linewidth]{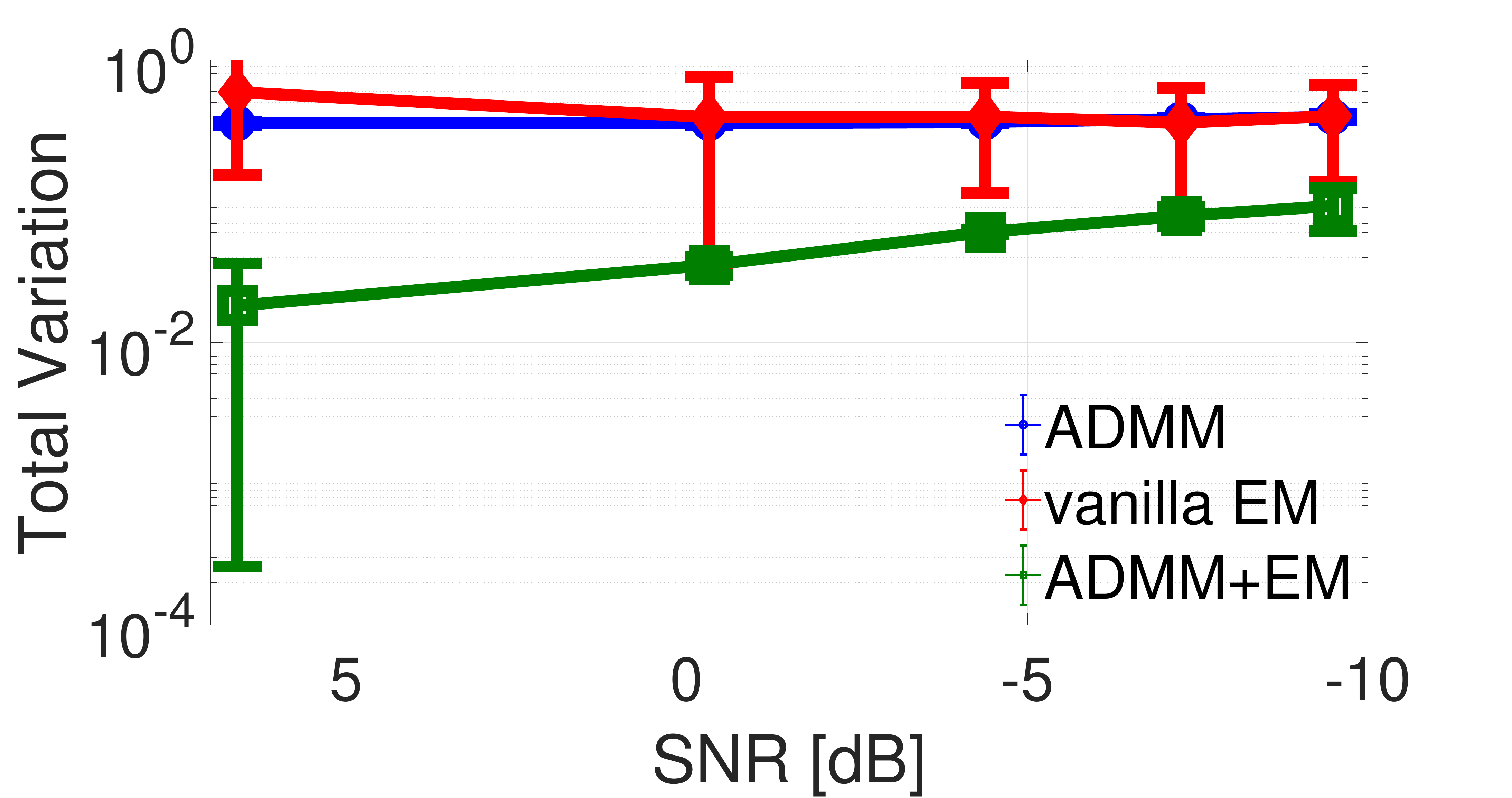}
		\label{fig:TV}
	}
	\vspace{-0.3cm}
	\caption{ \protect \subref{fig:RE} Relative error of the reconstruction, and \protect \subref{fig:TV} Total variation of the estimated view angle distribution. The means and the standard deviations are computed over 20 trials.  
	}
	\label{fig:TV and RE}
		\vspace{-0.6cm}
\end{figure}

Finally, we perform experiments on projection data at a wide range of noise levels to test the performance of our algorithms. The SNRs range from 6.61dB to -9.49dB. At each noise level, 20 independent trials are conducted and the initialization $\mathbf{a}^{(0)}$ and $\mathbf{p}^{(0)}$ are kept identical for all algorithms at each trial. The 2-D image is of size $64 \times 64$ pixels (see Fig.~\ref{fig:general_result} and the sample size is $N = 10^4$. The weight $\lambda_2$ in the augmented Lagrangian in~\eqref{cost_func:ADMM} is set to be $5$, and other parameters are the same as that in the first experiment. The average run-time are 265.41s (500 iterations) for ADMM, 682.74s (100 iterations) for EM and 364.52s (100 iterations ADMM + 50 iterations EM) for ADMM+EM. Several samples of the recovered images are shown in Fig.~\ref{fig:general_result}, and the average relative error of recovered object and total variation of estimated $\widetilde{\mathbf{p}}$ are shown in Fig.~\ref{fig:TV and RE}. The combined method that uses ADMM for ab initio reconstruction and EM for refinement outperforms the ADMM and EM. The success rates, defined as the percentage of trials that satisfy $\text{RE}(I,\widetilde{I})\le0.3$, of ADMM and ADMM+EM are much higher than that of EM (see Tab.~\ref{table:success_rate}). This indicates that ADMM and ADMM+EM are less likely to get stuck at bad local critical points compared to EM. 

\vspace{-0.2cm}

\section{Conclusion}
\vspace{-0.2cm}
\label{sec:4}
In this paper, we propose a new approach for recovering a 2-D object from its noisy projection tilts taken at unknown view angles. Instead of determining the view angle ordering of the projection tilts, the moment features are used to estimate the object and view angle distribution. This problem is first formulated as constrained weighted nonlinear least squares and solved by an ADMM algorithm. In addition, the algorithm provides an efficient initialization for EM. Via the numerical experiments, we show that ADMM can recover the 2-D object and view angle distribution exactly for the clean case and the proposed ADMM and ADMM+EM outperform EM in relative error and total variation for noisy data. 

\bibliographystyle{IEEEbib}
\bibliography{refs}

\end{document}